# Network Load Analysis and Provisioning of MapReduce Applications

Nikzad Babaii Rizvandi[1,2], Javid Taheri[1], Reza Moraveji[1,2], Albert Y. Zomaya [1]

[1] Center for Distributed and High Performance Computing, School of Information Technologies,
University of Sydney, Sydney, Australia

[2] Networked Systems Theme, National ICT Australia (NICTA), Australian Technology Park, Sydney, Australia

nikzad@it.usyd.edu.au
albert.zomaya | javid.taheri@sydney.edu.au
rmor1398@uni.sydney.edu.au

*Abstract*—In this paper, we study the dependency between MapReduce configuration parameters and network load of fixed-size MapReduce jobs during the shuffle phase; then we propose an analytical method to model this dependency. Our approach consists of three key phases: profiling, modeling, and prediction. In the first stage, an application is run several times with different sets of MapReduce configuration parameters (here number of map tasks and number of reduce tasks) to profile the network load of an application in the shuffle phase on a given cluster. Then, the relation between these parameters and the network load is modeled by multivariate linear regression. For evaluation, three applications (WordCount, Exim Mainlog parsing, and TeraSort) are utilized to evaluate our technique on a 5-node MapReduce private cluster.

## I.   INTRODUCTION

Recently, businesses have started adopting MapReduce as a popular distributed computing framework for processing large-scaled data in both public and private clouds; e.g., many Internet endeavors have already deployed MapReduce architecture to analyze their core businesses by mining their produced data. Consequently, application developers stand to benefit from understanding performance trade-offs in MapReduce-style computations by better utilizing their computational resources.

One common application in MapReduce is to repeat processing of fixed-size data. For example, system administrators are always interested to frequently analysis system log files (such as Exim MainLog files[2]). As these log files are captured with fix sampling rate, their size does not change from one month to another month. Another example is seismic imaging data where fix number of ultrasound senders/receivers produce earth underground information in a specific region; therefore, the size of output file (which is in order of terabyte) may slightly change from one experiment to another [3]. Another example is to find a sequence matching between a new RNA and RNAs in a database [4]. Generally, the size of Database (such as NCBI [5]) is almost the same in a large period of time. As these MapReduce applications –generally consume resources heavily– are repeated frequently, so it becomes important to improve their resource usage pattern.

Besides the simplicity of MapReduce framework, there are a few drawbacks. One major drawback is its heavy load on the cluster network during the map, shuffle and reduce phases (figure 1). The network load is of special concern with MapReduce as large amounts of traffic can be generated in the shuffle phase when the output of map tasks is transferred to reduce tasks. As each reduce task needs to read the output of all map tasks, a sudden explosion of network traffic can significantly deteriorate cloud performance. This is especially true when data has to traverse greater number of network hops while going across *racks* of servers in a data centre [6].

The technique in this paper is our attempt to study and model network load of MapReduce applications in their shuffle phase. For a given MapReduce platform, applications run iteratively with different values of two configuration parameters (number of map tasks, and number of reduce tasks) on fixed-size input data and network load in shuffle phase of these applications are gathered. Then for each application, a model is constructed by applying polynomial multi-linear regression on the set of configuration parameters values (as input) and obtained network loads of the application (as output).

This modeling, however, works under some assumptions. First, both complexity degree of an application and proper model selection influence the accuracy of modeling, resulting in less accuracy for high complex applications. Second, even though the modeling is valid for applications on different platforms, different MapReduce/Hadoop clusters should result in different model parameters.

The rest of this paper is organized as follows: section II highlights the related works in this area. Section III describes the dependency analysis and our analytical approach to profile, model and predict MapReduce applications' network load, followed by experimental results and conclusion in sections IV and V, respectively.

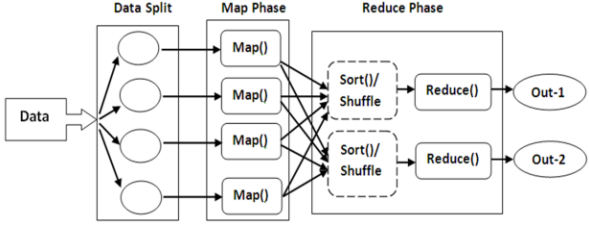

Figure 1. MapReduce workflow [1]

## II. RELATED WORKS

A statistics-driven workload modelling was introduced in [7] to effectively evaluate design decisions in scaling, configuration and scheduling. The framework in this work was utilized to make appropriate suggestions to improve the energy efficiency of MapReduce. Another modelling method was proposed in [9] for finding the total execution time of a MapReduce application. It used Kernel Canonical Correlation Analysis to obtain the correlation between the performance feature vectors extracted from MapReduce job logs, map time, reduce time, and total execution time. These features were acknowledged as critical characteristics for establishing any scheduling decisions. Recent works in [8, 9] reported a basic model for MapReduce computation utilizations. Here, at first, the map and reduce phases were modeled using dynamic linear programming independently; then, these phases were combined to build a global optimal strategy for MapReduce scheduling and resource allocation. In [1, 10] and later in [11], a Dynamic Time Warping based method was proposed to find the similarity between Cloud/MapReduce applications due to their CPU usage time series. In [6], a MapReduce resource allocation system was presented to enhance the performance of MapReduce jobs in the cloud by locating intermediate data to the local machines or close-by physical machines. This locality-awareness reduces network traffic in the shuffle phase generated in the cloud data center.

However, most works in resource provision and performance enhancement of MapReduce applications concentrate on CPU utilization of such applications and to the best of our knowledge there is no proper work on analysing network load of MapReduce applications and its relation to the configuration parameters such as number of map/reduce tasks. Generally, network of the MapReduce cluster is stressed during (1) the shuffle phase where each reducer contacts all other reducers –most probably on other machines in a cluster/cloud– to collect intermediate files, and (2) the reduce output phase where the final results of the whole job will be written to HDFS –usually with three replicas. Among them, the former is the most intensive period of network load – the focus of this paper – and acts as a performance issue in most MapReduce application. Therefore, from a cloud perspective, it would be interesting and useful to analysis and provision network load of a submitted application before its actual running.

## III. MODEL GENERATION AND EVALUATION

### A. Profiling

For each application, we carry out several experiments with different values of the number of map tasks and the number of reduce tasks on a given cluster. After running each experiment, the network load in the shuffling phase of an application is extracted (using SysStat API[12]) as training data for future use by the model. Due to the temporal changes, it is expected that several runs of an experiment –with the same configuration parameters– may result in slightly different network loads. Therefore, the average of several running of an experiment is considered as the network load.

### B. Model generation

In this section we explain how to model the relation between the configuration parameters and network load of an application in MapReduce. The problem of modeling based on multivariate linear regression involves choosing the suitable coefficients of the modeling such that the model's output well approximates a real system's response.

Consider three degree linear algebraic equations for M number of experiments of an application for $N$ effective configuration parameters ($M \gg N$) [13]:

$$\begin{cases} Net_{load}^{(1)} = \alpha_0 + \alpha_{11}p_1^{(1)} + \alpha_{12}(p_1^{(1)})^2 + \alpha_{13}(p_1^{(1)})^3 + \\ \quad \ldots + \alpha_{N1}p_N^{(1)} + \alpha_{N2}(p_N^{(1)})^2 + \alpha_{N3}(p_N^{(1)})^3 \\ Net_{load}^{(2)} = \alpha_0 + \alpha_{11}p_1^{(2)} + \alpha_{12}(p_1^{(2)})^2 + \alpha_{13}(p_1^{(2)})^3 + \\ \quad \ldots + \alpha_{N1}p_N^{(2)} + \alpha_{N2}(p_N^{(2)})^2 + \alpha_{N3}(p_N^{(2)})^3 \\ \quad \vdots \\ Net_{load}^{(M)} = \alpha_0 + \alpha_{11}p_1^{(M)} + \alpha_{12}(p_1^{(M)})^2 + \alpha_{13}(p_1^{(M)})^3 + \\ \quad \ldots + \alpha_{N1}p_N^{(M)} + \alpha_{N2}(p_N^{(M)})^2 + \alpha_{N3}(p_N^{(M)})^3 \end{cases}$$
(1)

where $Net_{load}^{(k)}$ is the value of network load during the shuffle phase of an application in the k-th experiment and $(p_1^{(k)}, p_2^{(k)}, \ldots, p_N^{(k)})$ are the values of N configuration parameters for the same experiment, respectively. With matrix P as:

$$P = \begin{bmatrix} 1, p_1^{(1)}, (p_1^{(1)})^2, (p_1^{(1)})^3, \ldots, p_N^{(1)}, (p_N^{(1)})^2, (p_N^{(1)})^3 \\ 1, p_1^{(2)}, (p_1^{(2)})^2, (p_1^{(2)})^3, \ldots, p_N^{(2)}, (p_N^{(2)})^2, (p_N^{(2)})^3 \\ \vdots \\ 1, p_1^{(M)}, (p_1^{(M)})^2, (p_1^{(M)})^3, \ldots, p_N^{(M)}, (p_N^{(M)})^2, (p_N^{(M)})^3 \end{bmatrix} \quad (2)$$

Eqn. (1) can be rewritten in the matrix format as:

$$\underbrace{\begin{bmatrix} Net_{load}^{(1)} \\ Net_{load}^{(2)} \\ \vdots \\ Net_{load}^{(M)} \end{bmatrix}}_{Net_{load}} = P \underbrace{\begin{bmatrix} \alpha_0 \\ \alpha_{11} \\ \alpha_{12} \\ \alpha_{13} \\ \vdots \\ \alpha_{N1} \\ \alpha_{N2} \\ \alpha_{N3} \end{bmatrix}}_{A} \quad (3)$$

Using the above formulation, the approximation problem is converted to estimating the values of model parameters, i.e. $\widehat{\alpha_0}, \widehat{\alpha_{11}}, \widehat{\alpha_{12}}, \widehat{\alpha_{13}} \ldots, \widehat{\alpha_{N1}}, \widehat{\alpha_{N2}}, \widehat{\alpha_{N3}}$, to optimize a cost function between the approximation and real values of the network load. Then, an approximated total network usage $\left(\widehat{Net_{load}}^{(*)}\right)$ of the application for a new unseen experiment is predicted as:

$$\widehat{Net_{load}}^{(*)} = \widehat{\alpha_0} + \widehat{\alpha_{11}} p_1^{(*)} + \widehat{\alpha_{12}} \left(p_1^{(*)}\right)^2 + \widehat{\alpha_{13}} \left(p_1^{(*)}\right)^3 + \cdots + \widehat{\alpha_{N1}} p_N^{(*)} + \widehat{\alpha_{N2}} \left(p_N^{(*)}\right)^2 + \widehat{\alpha_{N3}} \left(p_N^{(*)}\right)^3 \quad (4)$$

It can be mathematically proved that the model parameters can be calculated by minimizing least square error between real and approximated values as:

$$A = (P^T P)^{-1} P^T Net_{load} \quad (5)$$

*C. Evaluation Criteria*

We evaluate the accuracy of the fitted models, generated from regression based on a number of metrics [14]: Mean Absolute Percentage Error (MAPE), PRED(25), Root Mean Squared Error (RMSE) and R2 Prediction Accuracy. We describe the metrics in the following subsections.

- *Mean Absolute Percentage Error (MAPE)*

The Mean Absolute Percentage Error[14] for prediction model is given by the following formula:

$$MAPE = \frac{\sum_{i=1}^{M} \frac{\left|Net_{load}^{(i)} - \widehat{Net_{load}}^{(i)}\right|}{Net_{load}^{(i)}}}{M}$$

where $Net_{load}^{(i)}$ is the actual output of the application, $\widehat{Net_{load}}^{(i)}$ is the predicted output and $M$ is the number of observations in the dataset for which the prediction is made. A lower value of MAPE implies a better fit of the prediction model; i.e., indicating superior prediction accuracy.

- *PRED(25)*

The measure PRED(25)[14] is defined as the percentage of observations whose prediction accuracy falls within 25% of the actual value. A more formal definition of PRED(25) is as follows:

$$PRED(25) = \frac{\text{\# of observations with relative error less than } 25\%}{\text{\# of total observations}}$$

It is intuitive that a PRED(25) value closer to 1.0 indicates a better fit of the prediction model.

- *Root Mean Squared Error (RMSE)*

The metric Root Mean Square Error (RMSE)[14] is defined by the following formula:

$$RMSE = \sqrt{\frac{\sum_{i=1}^{M}(Net_{load}^{(i)} - \widehat{Net_{load}}^{(i)})^2}{M}}$$

A smaller RMSE value indicates a more effective prediction scheme.

- $R^2$ *Prediction Accuracy*

The $R^2$ Prediction Accuracy[14] is a measure of the goodness-of-fit of the prediction model. The formula of $R^2$ Prediction Accuracy is:

$$R^2 = 1 - \frac{\sum_{i=1}^{M}(Net_{load}^{(i)} - \widehat{Net_{load}}^{(i)})^2}{\sum_{i=1}^{M}(\widehat{Net_{load}}^{(i)} - \sum_{r=1}^{M} \frac{Net_{load}^{(r)}}{M})}$$

Note that, the $R^2$ value falls within the range [0, 1]. This metric is commonly applied to Linear Regression models. In fact, $R^2$ Prediction Accuracy determines how the fitted model approximates the real data points. A $R^2$ prediction accuracy of 1.0 indicates that the forecasting model is a perfect fit.

IV. EXPERIMENTAL RESULTS

*A. Experimental setting*

Three applications are used to evaluate the effectiveness of our proposed method. Our method has been implemented and evaluated on a 5-node physical MapReduce platform running Hadoop version 0.20.2 –Apache implementation of MapReduce developed in Java [15]. The hardware specification of the nodes in our 5-node MapReduce platform is:

- Master/node-0 and node-1: Dell with one processor: 2.9GHz, 32-bit, 1GB memory, 30GB Disk, and 512KB cache.
- Node-2, node-3, and node-4: Dell with one processor: 2.5GHz, 32-bit, 512MB memory, 60GB Disk, and 254KB cache.

These nodes were connected via LAN network links. In the training phase of our modeling, 64 *s*ets of experiments are conduced where the number of map/reduce tasks are a value in [4,8,12,16,20,24,28,32]; the size of input data is fixed to $12GB$. To overcome temporal changes, each experiment is repeated ten times. Then in the prediction phase, the accuracy of the application model is evaluated with 30 new/unseen experiments on the same input data size and random number of map/reduce tasks –as an integer value– in a range of $[4 \dots 32]$.

*TABLE 1. The prediction evaluation*

|  | RMSD | MAPE | $R^2$ prediction accuracy | PRED |
|---|---|---|---|---|
| WordCount | 0.24 | 1.78 | 0.93 | .93 |
| Exim MainLog parsing | 0.29 | 2.63 | 0.91 | .96 |
| TeraSort | 0.31 | 6.61 | 0.80 | 0.82 |

Our benchmark applications were WordCount (used by leading researchers in Intel [16], IBM [17], MIT [18], and UC-Berkeley [19]), TeraSort (as a standard benchmark international TeraByte sort competition [20, 21] as well as many researchers in IBM [22, 23], Intel [16], INRIA [24] and UC-Berkeley [25]), and Exim Mainlog parsing [1, 11]. These benchmarks were used due to their striking differences and also because other studies have relied on these benchmarks:

- *WordCount[11]:* This application reads data from a text file and, counts the frequency of each word. Results are written in another text file; each line of the output file contains a word and the number of its occurrence, separated by a TAB. In running a WordCount application on MapReduce, each mapper picks a line as input and breaks it into words $<key, value>$. Then it assign a $<key, value>$ pair to each word as $<word, 1>$. In the reduce stage, each reducer counts the values of pairs with the same $key$ and returns occurrence frequency (the number of occurrence) for each word,

- *TeraSort:* This application is a standard MapReduce sorting algorithm with a custom reducer –each reducer receives a sorted list of $N-1$ sampled $keys$ with predefined ranges. In particular, all $keys$ with $Sample[i-1] \leq key \leq Sample[i]$ are sent to $i^{th}$ reducer. This guarantees that the output of the $i^{th}$ reducer is always less than outputs of the $(i+1)^{th}$ reducer.

- *Exim MainLog parsing [2]*: Exim is a message transfer agent (MTA) for logging information of sent/received emails on Unix systems. This information that is saved in exim_mainlog files usually results in producing extremely large files in mail servers. To organize such massive amount of information, a MapReduce application is used to parse the data –in an exim_mainlog file– into individual transactions; each separated and arranged by a unique transaction ID.

*B. Results and Future work*

To test the accuracy of an application's model, we use it to predict network load of several experiments on different applications with random number of map/reduce tasks. We executed all experiments on a 5-node cluster and gathered their real network load to determine the prediction error. Figure 2 shows the prediction accuracies and MAPE prediction errors of these applications between actual values of network load as well as their predicted values. Table 1 shows the RMSD, MAPE, R2 prediction accuracy, and PRED of prediction for these applications. From this table, it can be seen that MAPE for WordCount and Exim MainLog parsing is in a reasonable margin (1.59 and 2.28, respectively) whereas it is slightly high for TeraSort (7.26). This implies that three-degree polynomial regression performs well for WordCount and Exim, while it almost fails to correctly model the network load of TeraSort; therefore a better model must be used for this application. Again, RMSD of both WordCount and Exim MainLog shows smaller values than that of TeraSort and experientially proves applicability of this modelling for these applications over TeraSort. This fact also supports according to PRED(25), where these two applications have more correct prediction rate than that of TeraSort.

In future, we plan to improve our prediction model by using the concept of model selection and also use other prediction methods like Neural Network. We also intend to use a variety of CPU and I/O intensive applications (such as, distributed Grep[11], Permutation Generator[6])). The type of application is important as an educated guess to explain the phenomenon in Table 1 could be related to the number of nodes in cluster that are utilized during the shuffling phase. For example, if an application is Reduce-input bound (like Permutation Generator), we should expect less data in shuffling phase and therefore fewer networks load in this phase. Shuffling phase needs to collect data from all nodes in the cluster and distribute it to all other nodes in Map and Reduce-input bound applications (like TeraSort). It is also worth noting that more nodes usually result in more system noise –such as I/O waiting, and fault in nodes– as well; this can significantly affect prediction accuracy of such systems.

V. CONCLUSION

This paper utilizes polynomial regression to model the dependency between two major MapReduce configuration parameters (number of map tasks and number of reducer tasks) and network load during the shuffle phase of

MapReduce applications with fixed-size input data. After extracting the network load of several experiments of an application with different values for the numbers of map/reduce tasks, multivariate regression is used to model the relation between the extracted network load and the used values for these two configuration parameters. Evaluation results on three applications on a 5-node MapReduce cluster show that our modeling technique can effectively predict the network load of these applications with root mean squared error of less than 7.5%.

## VI. ACKNOWLEDGMENT

The work reported in this paper is in part supported by National ICT Australia (NICTA). Professor A.Y. Zomaya's work is supported by an Australian Research Council Grant LP0884070.

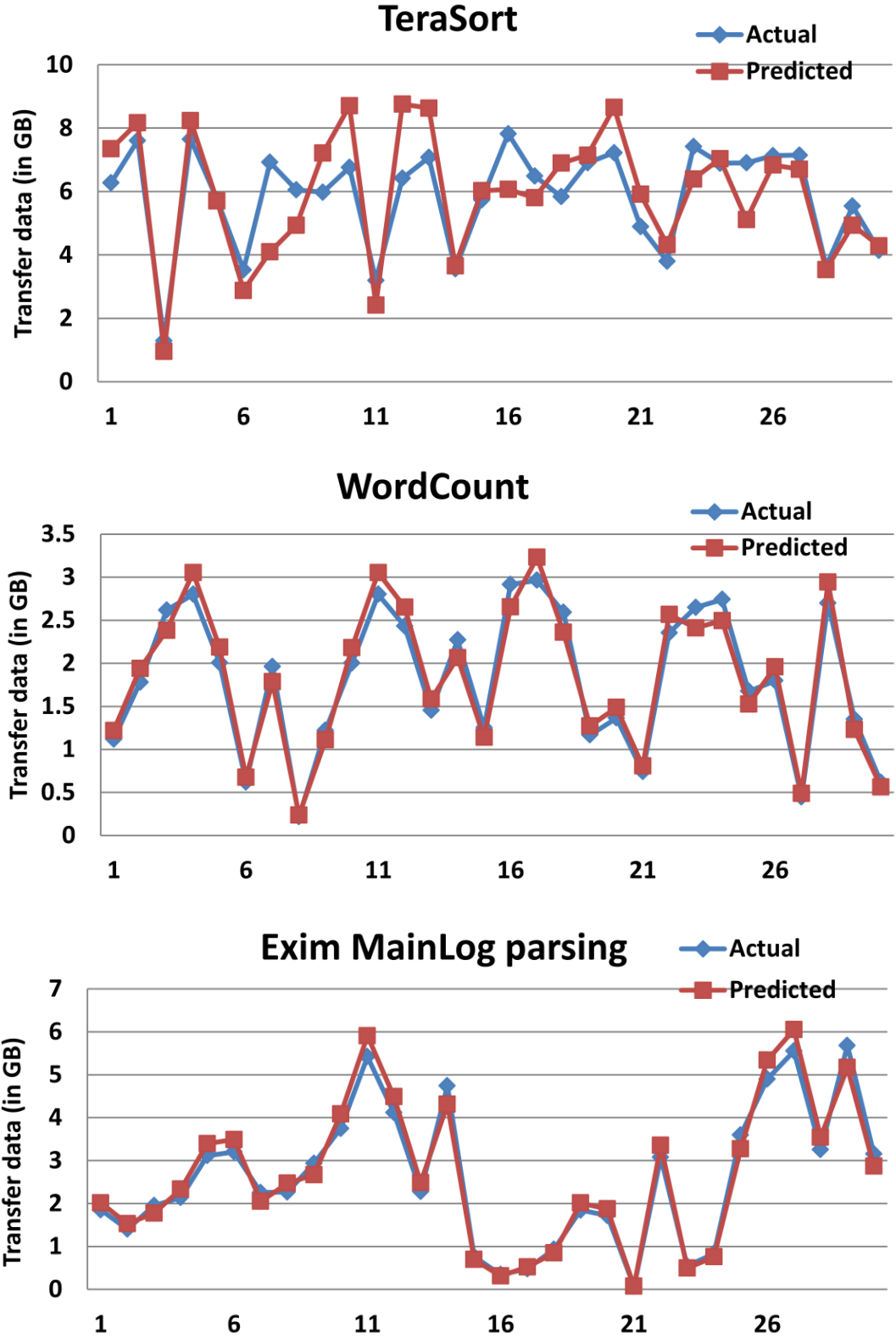

*F*igure 2. the prediction accuracy and error between the actual and predicted total network load for benchmark applications. The X-axis is the index of new/unseen experiments